\documentclass[aps,prb,twocolumn,superscriptaddress,english]{revtex4-1}
\pdfoutput=1

\usepackage{amsmath, amssymb, mathtools}
\usepackage{graphicx}% Include figure files
\usepackage{dcolumn}% Align table columns on decimal point
\usepackage{bm}% bold math
\usepackage{ifthen}
\usepackage{booktabs}
\usepackage{color}
\usepackage{verbatim}

\def\r{{\bf r}}

\def\k{{\bf k}}
\def\q{{\bf q}}

\def\S{\mathcal{S}}

\def\bra{\langle}
\def\ket{\rangle}

\def\f{\mathbf{f}}

\def\B{\mathbf{B}}

\def\m{\mathbf{m}}
\def\w{\omega}

\def\PSI{\overrightarrow{\psi}}
\def\U{\overrightarrow{u}}
\def\dU{\delta\overrightarrow{u}}

\begin{document}

\title{ 
\textit{Ab initio} calculation of spin fluctuation spectra using time dependent density functional 
perturbation theory, planewaves, and pseudopotentials}

\author{Kun Cao}
\affiliation{Department of Materials, University of Oxford, Parks Road, Oxford OX1 3PH,
United Kingdom}

\author{Henry Lambert \thanks{Current address: Physics Department, King’s College, London, United Kingdom}}
\affiliation{Department of Materials, University of Oxford, Parks Road, Oxford OX1 3PH,
United Kingdom}

\author{Paolo G. Radaelli}
\affiliation{Clarendon Laboratory, Department of Physics, University of Oxford, Parks Road,
Oxford OX1 3PU, United Kingdom}

\author{Feliciano Giustino}
\affiliation{Department of Materials, University of Oxford, Parks Road, Oxford OX1 3PH,
United Kingdom}

\date{\today }

\begin{abstract}
We present an implementation of time-dependent density functional perturbation theory for
spin fluctuations, based on planewaves and pseudopotentials. We compute the dynamic spin
susceptibility self-consistently by solving the time-dependent Sternheimer equation,
within the adiabatic local density approximation to the exchange and correlation kernel.
We demonstrate our implementation by calculating the spin susceptibility of representative 
elemental transition metals, namely bcc Fe, fcc Ni and bcc Cr. The calculated magnon 
dispersion relations of Fe and Ni are in agreement with previous work. 
The calculated spin susceptibility of Cr exhibits a soft-paramagnon
instability, indicating the tendency of the Cr spins to condense in a incommensurate spin 
density wave phase, in agreement with experiment.
\end{abstract}

\maketitle

\section{Introduction}\label{introduction}

Spin fluctuations play a central role in magnetic systems.\cite{moriya_book} For example they
underpin fundamental thermodynamic properties of magnets, such as the Curie temperature and
the heat capacity; they have long been discussed as a potential source of pairing in 
high-temperature superconductivity~\cite{scalapino12,monthoux07}; 
and they offer new opportunities in the development of spintronics~\cite{wolf01,zutic04,khitun10} 
and multiferroics.\cite{pimenov06}

In ordered magnets, spin fluctuations manifest themselves in two forms, magnons and Stoner 
excitations.\cite{kranendonk58, moriya_book} In standard textbooks, magnons or `spin waves' are typically 
illustrated as a collective, wave-like rotation of spins around their direction in the 
ground-state;\cite{kittel_book} Stoner or `spin flip' excitations correspond instead to electronic 
transitions from occupied to empty states, whereby the electron spin is reversed; this process
can alternatively be described as the creation of an electron-hole singlet pair of spins. 
When spin waves and spin flip excitations are degenerate, the spin wave is attenuated and sharp
magnon excitations cease to exist. This phenomenon is referred to as Landau damping,\cite{landau46}
and a clear discussion for the simplified case of the uniform electron gas can be found in 
Refs.~\onlinecite{moriya_book,buczek091}.
Landau damping is usually important in metallic systems,\cite{costa04} while it is generally
weak in insulators, owing to the large energy needed to excite Stoner pairs across the band gap.
 
The majority of studies on spin waves are currently based on the adiabatic approximation, whereby
spin and electronic excitations are decoupled.\cite{halilov98, gebauer00}
In this approach one maps the magnetic degrees of freedom into an array of local spins, and determines 
the coupling parameters from total energy calculations based on density functional theory (DFT), or from 
experiments. This approach proved to work well for insulating magnets, but carries some limitations. 
For example the adiabatic approximation does not admit Stoner excitations, therefore Landau damping is 
not captured, and the magnon energy renormalization resulting from Stoner excitations is 
absent.\cite{costa04,buczek091, buczek11} In addition this approach is sensitive to the choice of 
the discrete spin model and the determination of its parameters. It is clear that a more general 
approach is desirable in this context, thus motivating the need for a fully {\it ab initio} approach to spin waves.

The central quantity to describe spin fluctuations is the wave vector- ($\q$) and energy- ($\omega$) 
dependent spin susceptibility, $\chi(\q, \omega)$ (in the following Hartree atomic units will be
assumed). Magnon excitations correspond to the poles of $\chi(\q, \omega)$.\cite{moriya_book} 
A promising approach for calculating $\chi(\q, \omega)$ from first principles is given by
time-dependent density functional perturbation theory (TD-DFPT).\cite{gross85} The main appeal
of this method is that it allows one to describe spin waves and Stoner excitations on the
same footing, without invoking materials-specific approximations. TD-DFPT for spin fluctuations
has already been demonstrated using either all-electron\cite{savrasov98,buczek11} or 
pseudopotential\cite{rousseau12} implementations. Using the adiabatic local density approximation
(ALDA) to the exchange and correlation kernel,\cite{runge84,gross85,vonbarth72, callaway75}
it was shown that the formalism can capture the experimental magnon spectra of typical
transition metals (Fe, Co, Ni, Cr) with reasonable accuracy,\cite{savrasov98,buczek11,rousseau12}
provided the underlying DFT calculations could reproduce measured Stoner splittings.

Given these encouraging results on the use of TD-DFPT for calculating spin fluctuations, it would
be desirable to have these techniques available in the context of popular DFT implementations based
on planewaves and pseudopotentials. The first implementation of this type was reported not long 
ago.\cite{rousseau12} In this implementation the authors calculated the spin susceptibility using
the sum-over-states approach, and dealt with Brillouin zone sampling using maximally-localized
Wannier functions.~\cite{marzari12}

In this work we present a planewaves/pseudopotential implementation of TD-DFPT for spin fluctuations 
which does not rely on unoccupied electronic states. Our implementation employs the time-dependent
Sternheimer equation, in the spirit of related work in the area of GW calculations.
\cite{wilson08,wilson09,giustino10,umari10,umari11,nguyen12,lambert13} The implementation 
is based on the linear-response modules of the Quantum ESPRESSO materials simulation 
suite,\cite{giannozzi09} and is currently hosted on our GitHub repository.\cite{qelink} 
We demonstrate this development by calculating 
the spin fluctuations spectra of bcc iron, fcc nickel and bcc chromium. Our results are in
good agreement with previous calculations. The calculated spectra of Fe and Cr are in agreement
with experiment, while our results for Ni deviate from experiments owing to the incorrect
Stoner splitting in the underlying DFT calculation.

The manuscript is organized as follows. In Sec.~\ref{sec:theory} we present the TD-DFPT formalism 
for the spin susceptibility and the time-dependent Sternheimer equation. We also discuss how to
treat fractional occupations in metals, and how to perform the symmetry-reduction of the Brillouin
zone. At the end of this section we provide some technical details of the implementation.
In Sec.~\ref{sec:benchmark} we present calculation results on the elemental transition metals
Fe, Ni, and Cr. Here we compare our results to previous calculations as well as experiments.
We offer our conclusions in Sec.~\ref{sec:conclusion}, together with an outlook on future work.

\section{Spin susceptibility in time-dependent density functional perturbation theory}\label{sec:theory}

\subsection{General theory}\label{sec.dyson}
In this section we summarize the generalization of the TD-DFPT formalism\cite{gross85} 
to non-collinear spins, as already discussed in Refs.~\onlinecite{savrasov98,karlsson00,sasioglu10,
buczek11,rousseau12}. In the following we do not consider spin-orbit coupling, and we use the
spin $g$-factor $g=2$.
The DFT Kohn-Sham equations for a non-collinear spin system read:
  \begin{equation}
  \left[-\frac{\nabla^2}{2} \hat{I}+ \hat{V}_{\rm scf}({\bf r}) \right] \PSI_{n\k}({\bf r}) 
  = \epsilon_{n\k} \PSI_{n\k}({\bf r}),
  \label{KohnSham}
  \end{equation}
where $\hat{I}$ is the 2$\times$2 identify matrix, and $\hat{V}_{\rm scf}$ is the Kohn-Sham potential,
expressed as the following 2$\times$2 matrix:
  \begin{equation}
  \hat{V}_{\rm scf}({\bf r}) =V_{\rm scf}({\bf r}) \hat{I}
      +{\boldsymbol \sigma} \cdot {\bf B}_{\rm scf}({\bf r}).
  \label{eq:vks}
  \end{equation}
In this expression $V_{\rm scf}$ is the scalar part of the self-consistent potential, 
and $\B_{\rm scf}$ is the effective magnetic field arising from external potential as 
well as exchange and correlation.\cite{giustino_book} The Pauli matrix is 
given by ${\boldsymbol \sigma} = \sigma^1 {\bf u}_x + \sigma^2 {\bf u}_y +\sigma^3 {\bf u}_z$,
with $\sigma^i$ the usual 2$\times$2 Pauli matrices, and the unit vectors such as ${\bf u}_x$ 
denoting Cartesian directions.
$\PSI_{n\k}$ is a Kohn-Sham two-spinor eigenfunction with wavevector $\k$, band index $n$, and 
energy $\epsilon_{n\k}$, and corresponds to two scalar functions dependent on the space
coordinate ${\bf r}$, as follows:
  \begin{equation}
  \PSI_{n\k}({\bf r})= \begin{bmatrix} \psi_{n\k}^1({\bf r}) \\[3pt] \psi_{n\k}^2({\bf r}) \end{bmatrix}.
  \end{equation}
Using this notation, the 2$\times$2 density matrix becomes:
  \begin{equation}\label{eq.denmat}
  \rho_{\alpha \beta}({\bf r})=\frac{1}{N_\k}
  \sum_{\k,n\in{\rm occ}}
   \psi_{n\k}^{\alpha, *}({\bf r})\,\psi_{n\k}^{ \beta}({\bf r}) , \quad \alpha, \beta=1, 2,
  \end{equation}
where $N_\k$ is the number of $\k$ points used to discretize the first Brillouin zone (we assume
a uniform sampling), and the asterisc denotes complex conjugation. 
Using the Pauli matrix ${\boldsymbol \sigma}$, the density matrix can be decomposed as in 
Eq.~(\ref{eq:vks}): 
  \begin{equation}
  \hat{\rho}({\bf r})=\dfrac{1}{2}\left[n({\bf r}) \hat{I} + 
    {\boldsymbol \sigma} \cdot {\bf m}({\bf r})\right],
  \end{equation}
where $n(\r)$ is the standard electron charge density, and $\m(\r)$ is the electron spin density. 

In order to facilitate the algebra it is convenient to introduce an alternative 
expression for the density matrix, using vector notation:
  \begin{equation}
   \rho^0 = n, \,\,\rho^{1} = m_{x}, \,\,  \rho^{2} = m_{y},
    \,\,  \rho^{3} = m_{z}.
  \end{equation}
Using these definitions, together with $\sigma^0 = \hat{I}$ we can rewrite Eq.~(\ref{eq.denmat})
as follows:
  \begin{equation}
  \rho^i(\r) = \frac{1}{N_{\k}} \sum_{\k,n \in {\rm occ}} \PSI_{n\k}(\r)^{\dagger}\,
  \sigma^i\,\PSI_{n\k}(\r), \quad i=0,\dots 3,
  \end{equation}
where the dagger indicates the Hermitian conjugate.

We now consider a time-dependent external perturbation, which can be either an electric potential
or a magnetic field, $\delta V_{\rm ext}^i(\r t)$, with:
  \begin{equation}
  \delta V_{\rm ext}^0(\r t) = V_{\rm ext}(\r t), \quad \delta 
  V_{\rm ext}^{1}(\r t) = B_{\rm ext}^x(\r t),
  \end{equation}
and similarly for $j=2,3$. We neglect diamagnetic effects, so that the magnetic field
$\B_{\rm ext}$ only couples to the spin degrees of freedom.
In linear-response theory the variation of the density matrix in response to the external
perturbation is written as:
  \begin{equation}
  \label{eq:response}
  \delta \rho^i(\r t)={\sum}_j\int d(\r^\prime t')\,\chi^{ij}(\r t, \r^\prime t^\prime)\,
  \delta V^j_{\rm ext}(\r^\prime t'),
  \end{equation}
where the sum runs over the components of the four-vector, and $\chi^{ij}(\r t, \r^\prime t^\prime)$ 
is the generalized susceptibility. In TD-DFPT the generalized susceptibility is formally obtained
via a Dyson's equation:\cite{gross85} 
  \begin{eqnarray}
  \label{eq:dyson}
  \nonumber
  &&\chi^{ij}(\r t, \r^{\prime}t^\prime) = \chi^{ij}_{\rm KS}(\r t, \r^{\prime}t^\prime) 
  + \sum_{kl}\int \!d(\r_1t_1) d(\r_2 t_2) \\
  && \quad \times \chi^{ik}_{\rm KS}(\r t, \r_1 t_1) 
  \left[f_{xc}^{kl}(\r_1 t_1,\r_2 t_2)+\frac{2\delta_{k0}\delta_{l0}\delta(t_1-t_2)}{|\r_1-\r_2|}\right] 
  \nonumber \\
  && \quad \times \chi^{lj}(\r_2t_2, \r^{\prime}t^\prime),
  \end{eqnarray}
where $f_{xc}^{kl}$ is the exchange and correlation kernel, and
$\chi^{ij}_{\rm KS}$ is the non-interacting Kohn-Sham susceptibility.

The exchange-correlation kernel is usually written within the adiabatic local spin-density
approximation (ALSDA) to time-dependent DFT, meaning that one uses the static LSDA kernel
at equal times:\citep{callaway75, gross85,vonbarth72,buczek11}
  \begin{equation}
  \label{eq:xckernel}
  f_{xc}^{ij}(\r t,\r^{\prime}t^{\prime})=\frac{\delta^2 E_{xc}}{\delta 
  \rho^i(\r)\rho^j(\r^{\prime})}\delta(\r-\r^{\prime})\delta(t-t^{\prime}).
  \end{equation}
We note that this choice carries the additional implicit approximation that the ALSDA kernel 
is derived by considering an electron gas with collinear spins.

The Kohn-Sham susceptibility appearing in Eq.~(\ref{eq:dyson}) is defined so as to yield the 
density-matrix response to a variation of the self-consistent potential, in analogy with 
Eq.~(\ref{eq:response}):
  \begin{equation}
  \label{eq:KSresponse}
  \delta \rho^i(\r t)={\sum}_j\int d(\r^\prime t')\,\chi^{ij}_{\rm KS}(\r t, \r^\prime t^\prime)\,
  \delta V^j_{\rm scf}(\r^\prime t').
  \end{equation}
In view of practical calculations it is more convenient to work in the frequency domain rather than
in the time domain. Since the unperturbed Hamiltonian is time-independent, the susceptibility depends
only on the time difference $t- t'$. As a result, the last equation can be rewritten in the frequency
domain as follows:
  \begin{equation}
  \delta \rho^i(\r, \w)={\sum}_j\int d\r^\prime\,\chi^{ij}_{\rm KS}(\r , \r^\prime , \w)\,
    \delta V^j_{\rm scf}(\r^\prime, \w),
  \end{equation}
having defined:
  \begin{eqnarray}
  \delta\rho^{i}(\r, \w) & = & \int \!\!d t \,\delta \rho^{i}(\r t)\,e^{-i\w t},
  \end{eqnarray}
and similarly for $\delta V_{\rm scf}^{i}(\r, \w)$ and $\chi^{ij}_{\rm KS}(\r,\r^\prime,\w)$.
Equations~(\ref{eq:response})-(\ref{eq:xckernel}) are transformed into the freqency domain along
the same lines. Using standard perturbation theory, the Kohn-Sham susceptibility in the frequency 
domain can now be written explicitly in terms of unperturbed Kohn-Sham spinors:~\cite{gross85}
  \begin{eqnarray}
  \nonumber
  &&\chi^{ij}_{\rm KS}(\r,\r^\prime, \w) = \frac{1}{N_\k^2}
  \sum_{nm,\k,\q} 
  \frac{f_{n\k}-f_{m\k+\q}}{\epsilon_{n\k} -\epsilon_{m\k+\q}+\omega} \nonumber \\[3pt]
  &&\qquad \times \PSI_{n\k}^{\dagger}(\r)\,\sigma^i\, \PSI_{m\k+\q}(\r)\,
  \PSI_{m\k+\q}^{\dagger}(\r^{\prime})\,\sigma^j\, \PSI_{n\k}(\r^{\prime}),\quad\,\,
  \label{eq:chiqKS}
  \end{eqnarray}
where $f_{n\k}$ and $f_{m\k+\q}$ are occupation numbers. 

This formulation provides a two-step procedure for calculating the generalized susceptibility. 
First, the Kohn-Sham susceptibility is calculated via Eq.~(\ref{eq:chiqKS}), starting from the
unperturbed Kohn-Sham spinor wavefunctions. The computed $\chi^{ij}_{\rm KS}$ is then used inside
Eq.~(\ref{eq:dyson}) (after transforming to the frequency domain), so as as to evaluate $\chi^{ij}$
and obtain spin fluctuation spectra.

The main disadvantage of this procedure is that the evaluation of $\chi_{\rm KS}$ via Eq.~(\ref{eq:chiqKS}) 
relies on the calculation of unoccupied Kohn-Sham states, and the convergence of the results with
respect to the number of empty bands is slow. In order to circumvent this bottleneck, here we employ 
an alternative approach which is based on the Sternheimer equation, \cite{sternheimer54}
and which requires one to compute only occupied Kohn-Sham states.

\subsection{Calculation of the spin susceptibility using the Sternheimer equation}

The Sternheimer equation in time-dependent perturbation theory reads:\cite{savrasov98}
  \begin{equation}
  \left(\hat{H}-i\frac{\partial}{\partial t}\hat{I}\right) \,\delta \overrightarrow{\psi}_{n\k} ({\bf r},t)
   = -(1-\hat{P}_{\rm occ})\,\delta \hat{V}_{\rm scf}(\r, t)\overrightarrow{\psi}_{n\k} ({\bf r}),
  \label{eq:sternheimer}
  \end{equation}
where $\hat{H}$ is the unperturbed Kohn-Sham Hamiltonian, corresponding to the term in 
square brackets in Eq.~(\ref{KohnSham}), $\delta \overrightarrow{\psi}_{n\k}$
is the first-order change of the spinor wavefunction, and $\delta \hat{V}_{\rm scf}$ is 
the first-order variation of the Kohn-Sham potential from Eq.~(\ref{eq:vks}). The operator
$\hat{P}_{\rm occ}$ is the projector on the manifold of unoccupied Kohn-Sham states.

Equation~(\ref{eq:sternheimer}) is naturally found as an intermediate step of the perturbation theory
approach; by expressing $\delta \overrightarrow{\psi}_{n\k}$ in the frequency domain and expanding
on a basis of (occupied and empty) Kohn-Sham states, this expression leads immediately to 
Eq.~(\ref{eq:chiqKS}). In order to avoid the computation of unoccupied states, Eq.~(\ref{eq:sternheimer})
must be solved directly as a non-homogeneous linear system.

The Sternheimer equation was originally introduced for calculating the static dielectric
polarizability of atoms,\cite{sternheimer54} and found widespread use in the DFT community
to compute phonon dispersion relations using DFPT.\cite{baroni01, savrasov92, gonze95}
The time-dependent version of the Sternheimer equation was used to calculate spin 
susceptibilities,\cite{savrasov98} dynamic molecular polarizabilities,\cite{andrade07} 
and the screened Coulomb interaction in GW calculations.\cite{reining97,giustino10,umari10,umari11, nguyen12,
lambert13}

We now consider an external perturbation given by a monochromatic planewave:
  \begin{equation}
  \label{eq:vext}
  \delta V_{\rm ext}^j ({\bf r}t)=\delta V_{\rm ext}^j(\q,
  \omega) \left[e^{i(\q\cdot\r+\omega t)-\eta t} + {\rm c.c.}\right],  
  \end{equation}
with $\eta$ being a positive infinitesimal.
This perturbation induces a variation of the density matrix that can be evaluated
using Eq.~(\ref{eq:sternheimer}). After some lengthy but otherwise straightforward
algebra one finds:
  \begin{eqnarray}
  \label{eq:responsep}
  \delta \rho^{i}_{\q}(\r,\omega) & = & \frac{1}{N_\k}\sum_{n\in{\rm occ},\k}\U_{n\k}^\dagger(\r)\,\sigma^i\,
  \delta\overrightarrow{u}_{n\k+\q}(\r,\omega) \nonumber \\
  &  +&\frac{1}{N_\k}\sum_{n\in\rm {occ},\k}\delta \U_{n\k-\q}^\dagger(\r,-\omega)\,\sigma^i\,\U_{\!n\k}(\r),
  \quad
  \label{eq:responsem}
  \end{eqnarray}
where $\U_{\!n\k}$ is the Bloch-periodic part of the Kohn-Sham wavefunction, i.e.
$\overrightarrow{\psi}_{n\k}(\r) = e^{i\k\cdot\r} \overrightarrow{u}_{\!n\k}(\r)$,
and similarly for $\delta\overrightarrow{u}_{\!n\k+\q}(\r,\omega)$ and $\delta \rho^{i}_{\q}(\r,\omega)$.
The variation of the Kohn-Sham wavefunctions appearing in the last expression can be
found from Eq.~(\ref{eq:sternheimer}), following the usual decoupling procedure employed
in DFPT for phonons:\cite{baroni01}
  \begin{eqnarray}
  \label{eq:blochSHp}
  & &(\hat{H}_{\k+\q}-\epsilon_{n\k} +\omega+i\eta)  \,\delta\overrightarrow{u}_{n\k +   
  \q}({\bf r}, +\omega) \nonumber \\[3pt]
  & &\qquad\qquad = -(1-\hat{P}_{\rm occ}^{\,\k +\q})\, \delta \hat{V}_{\rm scf}^{+\q}({\bf r},+ 
  \omega)\overrightarrow{u}_{\!n\k}({\bf r}), \\[4pt]
  \label{eq:blochSHm}
  & &(\hat{H}_{\k - \q}-\epsilon_{n\k} - \omega+i\eta) \,\delta \overrightarrow{u}_{n\k 
  - \q}({\bf r}, -\omega) \nonumber \\[3pt]
  & & \qquad\qquad = -(1-\hat{P}_{\rm occ}^{\,\k-\q})\, \delta \hat{V}_{\rm scf}^{- \q}({\bf r},
  -\omega)\overrightarrow{u}_{\!n\k}({\bf r}).
  \end{eqnarray}
Here $\hat{H}_{\k}$ denotes the ${\k}$-projected unperturbed Kohn-Sham Hamiltonian.
Similarly $\hat{P}_{\rm occ}^{\k}$ indicates the component of the projector operator
on the occupied states with wavevector ${\k}$.
$\delta \hat{V}_{\rm scf}^{+\q}(\r, \w)$ is the Bloch-periodic component of the self-consistent
variation of the Kohn-Sham potential, for the wavevector $+\q$.

The variation of the self-consistent potential is related to the variation of the
charge density as follows:
  \begin{eqnarray}
  \delta \hat{V}^{\q}_{\rm scf}({\bf r},\omega)& = & \sigma^j\,\delta V_{\rm ext}^j(\q, \w) \nonumber \\
  & + &\sigma^0 \!\int \frac{\delta \rho^{0}_{\q}
  (\r^\prime, \omega)}{|\r-\r^\prime|}e^{-i\q\cdot(\r-\r^\prime)}d\r^\prime \nonumber  \nonumber \\
  & + &\sum_{ij}\sigma^i \, f_{xc}^{ij}[{\rho(\r), \r}]\,
  \delta \rho^{j}_{\q}(\r, \omega),  \label{eq:dvks}
  \end{eqnarray}
where the integration in the second line is in the unit cell, and the exchange-correlation
kernel in the third line is evaluated for the unperturbed density.
From Eq.~(\ref{eq:blochSHm}) we see that, in addition to $\delta \hat{V}^{\q}_{\rm scf}({\bf r},\omega)$,
we also need $\delta \hat{V}_{\rm scf}^{- \q}({\bf r}, -\omega)$. In order to evaluate
this term via Eq.~(\ref{eq:dvks}) we simply change the signs of $\q$ and $\omega$, and observe
that $\delta \rho^{i}_{-\q}(\r,-\omega) =  \delta \rho^{i*}_{\q}(\r,\omega)$.

Equations~(\ref{eq:responsep})-(\ref{eq:dvks}) are to be solved self-consistently. The procedure
starts from Eq.~(\ref{eq:dvks}), by setting the initial variation of the Kohn-Sham potential
equal to the external perturbation (i.e. by retaining only the first line). Then Eqs.~(\ref{eq:blochSHp})
and (\ref{eq:blochSHm}) are solved, and the solutions are used in Eq.~(\ref{eq:responsep}) to obtain
$\delta \rho^{i}_{\q}(\r,\omega)$. We emphasize that the presence of an external magnetic field 
breaks time reversal symmetry, therefore  Eqs.~(\ref{eq:blochSHp}) and (\ref{eq:blochSHm})
have to be solved {\it separately}. This is in contrast to what happens in DFPT for phonons,
whereby these two equations become equivalent after taking into account time-reversal symmetry.\cite{baroni01}

The self-consistent solution of Eqs.~(\ref{eq:responsep})-(\ref{eq:dvks}) yields the variation
$\delta \rho^{j}_{\q}(\r, \omega)$ of the density matrix in response to the external perturnation
of Eq.~(\ref{eq:vext}). By taking the unit-cell average of the density variation,
$\delta \rho^i(\q, \omega) = \int \!d\r \,\delta \rho^{j}_{\q}(\r, \omega)$, we can finally
calculate the susceptibility as:
  \begin{equation}
  \chi^{ij}(\q, \omega)=\frac{\delta \rho^i(\q, \omega)}{\delta 
  V_{\rm ext}^j(\q, \omega)}.
  \end{equation}
This quantity can directly be compared with experiments. For example, 
in a ferromagnet with the spin density polarized along the $z$-direction, the transverse component 
of spin susceptibilty, defined as $\chi^{+-} = \chi^{11} - i\chi^{12}$, yields the inelastic 
neutron scattering cross section according to the equation:\cite{izuyama63}
  \begin{equation}
  \frac{\partial^2 \Sigma}{\partial\Omega \partial\omega} \propto {\rm Im}\, \chi^{+-}(\q, \w),
  \end{equation}
where $\q$ and $\omega$ have the meaning of momentum and energy transfer, respectively,
and $\partial\Omega$ is the element of solid angle spanned by $\q$.
Sharp peaks of ${\rm Im}\chi^{+-}$ in the $(\q, \w)$ plane correspond to magnon excitations.
\cite{moriya_book}

Magnon excitation are expected to be damped in the presence of resonant Stoner spin-flip excitations.
The region in the $(\q, \w)$ plane where Stoner excitations are allowed corresponds to energy and
momenta for which the non-interacting susceptibility, ${\rm Im}\,\chi_{\rm KS}^{+-}(\q, \w)$, is 
nonzero.\cite{moriya_book} In fact, by performing a Fourier transform of Eq.~(\ref{eq:chiqKS}), 
we find immediately:
  \begin{eqnarray}
  && {\rm Im}\, \chi_{\rm KS}^{+-}(\q,\w)= 
  \frac{\pi}{2} \frac{1}{N_\k}\sum_{nm,\k}(f_{n\k}-f_{m\k+\q}) \nonumber \\
   && \qquad\quad\times \left|\bra \U_{\!m\k+\q}| \sigma^- | \U_{\!n\k} \ket \right |^2 
    \delta(\epsilon_{n\k}-\epsilon_{m\k+\q}+\w),\quad
  \end{eqnarray}
where $\sigma^-=\sigma^1-i\sigma^2$. The r.h.s. of this expression corresponds to the standard
transition rate as given by the Fermi golden rule, with respect to an operator which lowers
the spin quantum number in a ferromagnet.

\subsection{Fractional occupations}

The formalism described in the previous section is applicable to insulators and semiconductors,
where the occupied and unoccupied states are separated by a finite energy gap. In principle, the 
formalism also applies for metals at zero temperature. However, in the case of metals, a very
dense sampling of the Brillouin zone would be required to correctly describe the Fermi surface.
To avoid this complication in the case of metals, it is common to perform Brillouin zone
integrals using the tetrahedron method,\cite{quong92,savrasov92, kawamura14} or to employ smearing 
techniques.\cite{gironcoli95,baroni01}
In this work we opted for the use of electronic smearing, and in the following we discuss how the
formalism introduced in Ref.~\onlinecite{gironcoli95} needs to be adapted to deal with
frequency-dependent perturbations and spinor wavefunctions.

In the scheme of Ref.~\onlinecite{gironcoli95} each Kohn-Sham energy level is broadened by 
a smearing function defined by $\delta_\gamma(\epsilon)=\tilde{\delta}(\epsilon/\gamma)/\gamma$.
Here $\tilde{\delta}(x)$ is a normalized function such that $\delta_\gamma(x)$ tends 
to the Dirac $\delta$ function when the smearing width $\gamma$ tends to zero. The simplest
smearing function is a Gaussian but there are many practical alternatives.\cite{giannozzi09}
From the definition of $\delta_\gamma$ one naturally obtains a smooth approximation to the step function,
$\tilde{\theta(x)}=\int^{x}_{-\infty}\tilde{\delta}(x')dx'$. 

Following Ref.~\onlinecite{gironcoli95}
we define $\tilde{\theta}_{n, m} = \tilde{\theta}[(\epsilon_n-\epsilon_m)/\gamma]$,
and  $\tilde{\theta}_{{\rm F}, m} = \tilde{\theta}[(\epsilon_{\rm F}-\epsilon_m)/\gamma]$, 
with $\epsilon_{\rm F}$ the Fermi energy. Using these definitions the smeared density matrix reads:
  \begin{equation}
  \rho^i(\r)=\frac{1}{N_\k}\sum_{n\k}\tilde{\theta}_{{\rm F},n\k}\,
  \U_{\!n\k}^{\dagger}(\r)\,\sigma^i \, \U_{\!n\k}(\r),
  \label{eq:rho}
  \end{equation}
where the sum is over all states $n$. Since $\tilde{\theta}$ is a smeared step function,
it is sufficient to only calculate all occupied states and a handfulf of unoccupied states,
up to the energy $\sim \epsilon_{\rm F} + 10 \gamma$.
The linear density response to an external monochromatic perturbation as in Eq.~(\ref{eq:vext})
reads:
  \begin{eqnarray}
  \label{eq:responsep1}
  \delta \rho^{i}_{\q}(\r,\omega) & = & \frac{1}{N_\k}\sum_{n\k}\tilde{\theta}_{F,n\k}
  \,\U_{n\k}^\dagger(\r)\,\sigma^i\,\delta\overrightarrow{u}_{\!\!n\k+\q}(\r,\omega) \nonumber \\
  &+ & \frac{1}{N_\k} \sum_{n\k}\tilde{\theta}_{F,n\k}\,\delta\overrightarrow{u}_{\!\!n\k-\q}^\dagger(\r,-\omega)
  \,\sigma^i\,\overrightarrow{u}_{\!\!n\k}(\r).\qquad
  \end{eqnarray}
In this expression we are neglecting the contribution to the density variation arising from
a change in the Fermi level, which corresponds formally to the variation of the prefactor
$\tilde{\theta}_{F,n\k}$. This contribution is only important for $\q=0$, and it vanishes for the perturbations
considered in this work.

The linear variation of the spinor wavefunction can be formally written using perturbation theory:
  \begin{equation}\label{eq:PSIp}
  \dU_{\!n\k+\q}(\r, \omega) 
  = \sum_m\frac{\U_{m\k+\q} \bra \U_{\!m\k+\q}|\delta \hat{V}_{\rm scf}^{\q}(\r, \omega)|\U_{\!n\k} \ket}
  {\epsilon_{n\k}-\epsilon_{m\k+\q}+\omega-i\eta},
  \end{equation}
and similarly for $\dU_{\!n\k-\q}(\r, -\omega)$. After replacing this expression in Eq.~(\ref{eq:responsep1})
one obtains:
  \begin{eqnarray}
  \nonumber &&
  \delta \rho^{i}_{\q}(\r,\w) =\frac{1}{N_\k}\sum_{nm,\k} (\tilde{\theta}_{F,n\k}-\tilde{\theta}_{F,m\k+\q}) 
   \\ 
  &&\qquad\times \frac{\U_{n\k}^\dagger\sigma^i\U_{m\k+\q} 
  \bra \U_{m\k+\q}|\delta \hat{V}_{\rm scf}^{\q}(\r, \omega)|\U_{n\k} 
  \ket}{\epsilon_{n\k}-\epsilon_{m\k+\q}+\omega-i\eta}.\qquad
  \label{eq:charge} 
  \end{eqnarray} 
To obtain a more compact expression, in the second line of Eq.~(\ref{eq:responsep1}) 
we replaced $\k$ by $\k+\q$ and $m$ by $n$, and we used the identity 
$\delta \hat{V}_{\rm scf}^{-\q}(\r, -\omega)^\dagger= \delta \hat{V}_{\rm scf}^{\q}(\r, \omega)$.

Equation~(\ref{eq:charge}) contains a summation over all states, occupied and empty. In order to
recast this expression into a sum over occupied states only, we observe that the prefactor
$\tilde{\theta}_{F,n\k}-\tilde{\theta}_{F,m\k+\q}$ vanishes when the states $n\k$ and $m\k+\q$
are both occupied or both unoccupied. This observation can be used to bring Eq.~(\ref{eq:charge})
into a form similar to Eq.~(\ref{eq:responsep1}), but without the $\tilde{\theta}$ prefactors.
To this aim we note that $\tilde{\theta}(x)+\tilde{\theta}(-x)=1$, therefore 
Eq.~(\ref{eq:charge}) can be rewritten as:
  \begin{widetext}
  \begin{eqnarray}
  \nonumber
  \delta \rho^{i}_{\q}(\r,\omega) & = &
  \frac{1}{N_\k}\sum_{nm,\k}\,\,\,\,(\tilde{\theta}_{F,n\k}-\tilde{\theta}_{F,m\k+\q})\tilde{\theta}_{m\k+\q,n\k}
  \frac{\U_{n\k}^\dagger\sigma^i\U_{m\k+\q} \bra \U_{m\k+\q}|\delta 
  \hat{V}_{\rm scf}^{\phantom{-}\q}(\r, \phantom{-}\omega)|\U_{n\k}\ket}
  {\epsilon_{n\k}-\epsilon_{m\k+\q}+\omega-i\eta} \\ \label{eq:final}
  & + &\frac{1}{N_\k} \sum_{nm,\k} \left[(\tilde{\theta}_{F,n\k}
  -\tilde{\theta}_{F,m\k-\q})\tilde{\theta}_{m\k-\q,n\k}
  \frac{\U_{n\k}^\dagger\sigma^i\U_{m\k-\q} \bra \U_{m\k-\q}|\delta 
  \hat{V}_{\rm scf}^{-\q}(\r, -\omega)|\U_{n\k} \ket}
  {\epsilon_{n\k}-\epsilon_{m\k-\q}-\omega-i\eta} \right]^\dagger.
  \end{eqnarray}
  \end{widetext}
Also in this case the second line has been rewritten by exchanging $\k$ and $\k-\q$, $n$ and $m$,
and using $\delta \hat{V}_{\rm scf}^{-\q}(\r, -\omega)^\dagger= \delta \hat{V}_{\rm scf}^{\q}(\r, \omega)$.
By inspecting the terms $(\tilde{\theta}_{F,n\k}-\tilde{\theta}_{F,m\k+\q})\tilde{\theta}_{m\k+\q,n\k}$
we can see that now the summation over $n$ effectively runs over occupied states, and that over $m$ runs over
unoccupied states. At this point it is a matter of alebra to show that the density matrix variation
can be written compactly as follows:
  \begin{eqnarray}
  \label{eq:responsep-new}
  \delta \rho^{i}_{\q}(\r,\omega) & = & \frac{1}{N_\k}\sum_{n\k}
  \U_{n\k}^\dagger(\r)\,\sigma^i\,\delta\overrightarrow{v}_{\!\!n\k+\q}(\r,\omega) \nonumber \\
  &+ & \frac{1}{N_\k} \sum_{n\k}\delta\overrightarrow{v}_{\!\!n\k-\q}^\dagger(\r,-\omega)
  \,\sigma^i\,\overrightarrow{u}_{\!\!n\k}(\r),\qquad
  \end{eqnarray}
where $\delta\overrightarrow{v}_{\!\!n\k+\q}(\r,\omega)$ is the solution of the {\it modified}
Sternheimer equation:
  \begin{eqnarray}\label{eq.stern2}
  & &(\hat{H}_{\k+\q}+\alpha\hat{Q}_{\k+\q}-\epsilon_{n\k} -\omega+i\eta)  \,\delta\overrightarrow{v}_{n\k +
  \q}({\bf r}, +\omega) \nonumber \\[3pt]
  & &\quad\qquad = -(\tilde{\theta}_{F,n\k}
  -\hat{P}^{+\omega}_{n,\k +\q})\, \delta \hat{V}_{\rm scf}^{+\q}({\bf r},+
  \omega)\overrightarrow{u}_{\!n\k}({\bf r}). \quad
  \end{eqnarray}
In this equation $\alpha$ is a real parameter to be discussed below, and the
projector operators are defined as:  
  \begin{eqnarray}
  \hat{Q}_{\k+\q} &=& {\sum}_m |\U_{m\k+\q} \ket \bra \U_{m\k+\q} |, \\
  \hat{P}^{+\omega}_{n,\k +\q} &=& {\sum}_m\beta_{nm,\k +\q}^{+\omega} 
  |\U_{m\k+\q} \ket \bra \U_{m\k+\q} |, \label{eq:scf2}
  \end{eqnarray}
with the summation running over the occupied states plus a few empty states, as discusses for
Eq.~(\ref{eq:rho}). The parameters $\beta_{nm,\k +\q}^+$ are given by:
  \begin{eqnarray}
  \beta_{nm,\k +\q}^{+\omega} & = &\tilde{\theta}_{F,n\k}\,\tilde{\theta}_{n\k, m\k+\q} 
  + \tilde{\theta}_{F,m\k+\q}\,\tilde{\theta}_{m\k+\q, n\k} \nonumber \\
  &+& \alpha \frac{(\tilde{\theta}_{F,n\k}-\tilde{\theta}_{F,m\k+\q})\,\tilde{\theta}_{m\k+\q,n\k}}
  {\epsilon_{n\k}-\epsilon_{m\k+\q}+\omega-i\eta}.\label{eq.beta}
  \label{eq:scf3}
  \end{eqnarray}
An equation analogous to Eq.~(\ref{eq.stern2}) is obained for the function 
$\delta\overrightarrow{v}_{\!\!n\k-\q}(\r,-\omega)$ needed in Eq.~(\ref{eq:responsep-new}). 
In practice one only needs to change the signs of $\q$ and $\omega$ in Eqs.~(\ref{eq.stern2})-(\ref{eq.beta}).
In order to derive Eqs.~(\ref{eq:responsep-new})-(\ref{eq.beta}) starting from Eq.~(\ref{eq:final})
it is sufficient to act on the first line of the r.h.s. with the operator 
$\epsilon_{n\k}-(\hat{H}_{\k+\q}+\alpha\hat{Q}_{\k+\q})+\omega-i\eta$.
These equations constitute a straightforward generalization of the treatment of fractional
occupations in standard DFPT for phonons.\cite{gironcoli95,baroni01}

As in the case of phonon calculations, the parameter $\alpha$ appearing in Eq.~(\ref{eq.stern2})
is chosen so as to make the linear operator on the l.h.s. non-singular. When choosing $\eta=0$,
the system can become singular if $\epsilon_{m\k+\q}$ is in resonance with $\epsilon_{n\k}+\omega$,
where $n$ belongs to the manifold of occupied states. To avoid a singularity it therefore suffices 
to choose $\alpha = \epsilon_F-\epsilon_{\rm min}+ \omega_{\rm max} +10\gamma$, where $\epsilon_{\rm min}$
is the smallest eigenvalue of the occupied manifold, and $\omega_{\rm max}$ is the highest frequency
considered in the calculations. If we set instead $\eta>0$, the calculation is effectively
performed for a complex frequency, and strictly speaking the linear system cannot become singular.
However, the use of the projector $\alpha\hat{Q}_{\k+\q}$ is still important in order to
reduce the condition number of the linear system.

\subsection{Symmetry reduction}

The summation over $\k$-points in the Brillouin zone in Eq.~(\ref{eq:responsep-new}) can 
effectively be performed by exploting crystal symmetry operations. In standard DFPT 
symmetry is used to reduce the set of $\k$-points to a symmetry-irreducible wedge of the Brillouin 
zone.\cite{giannozzi09} 

In the present work we are dealing with DFPT in the presence of spinor wavefunctions, therefore
we also need to take into account the effect of symmetry operations on the electron spins.
To this aim we make use of spin-space groups (SSG).\cite{sandratskii91} The action of an element $\S$ of
the SSG on a spinor wavefuntion can be described as:
  \begin{equation}
  \S: \{R_s|R|f\}\PSI(\r)=R_s\PSI(R^{-1}\r-\f),
  \end{equation}
where  $R$ is the spacial rotation, $\f$ is the possible fractional translation, and
$R_s$ is the matrix that rotates the spins.

Generally speaking one could also consider time-reversal symmetry in order to perform
further reductions of the number of required $\k$-points. Time-reversal can be combined
with operations in the SSG.\cite{sandratskii91} However, since in our present work we are focusing
on perturbations corresponding to magnetic fields,
time-reversal symmetry is broken. As a result we can only consider symmetry operations
in the SSG which do not involve time-reversal. In practice
we perform symmetry-reduction by considering all the symmetry operations $\S$ which do not
contain time-reversal and which belong
to the small group of $\q$, where $\q$ is the wavevector of the perturbation 
(the small group is the sub-group of elements which leave $\q$ unchanged, $\S\q=\q$).

\subsection{Implementation details}\label{sec:implementation}

The method described in the preceeding sections was implemented using planewaves basis sets 
and pseudopotentials, starting from the linear response modules of the Quantum ESPRESSO suite,
and in particular from the PHONON code.\cite{giannozzi09} 
The development version of this code is hosted on our GitHub repository \cite{qelink}. We have support for both
norm-conserving \cite{hamann79} and ultrasoft \cite{vanderbilt90} pseudopotentials.

In our implementation the Sternheimer equation, Eq.~(\ref{eq:responsep-new}), is solved separately
for each frequency $\omega$ using the complex biconjugate gradient method, as described in 
Ref.~\onlinecite{jacobs86}. The implementation was adapted from related work within the context
of GW calculations from Ref.~\onlinecite{giustino10,lambert13}.

In order to minimize fluctuations of the density matrix during the self-consistent iterations,
we employed a generalization of the modified Broyden method for charge-density mixing,\cite{johnson88}
following Ref.~\onlinecite{giustino10}. We found that typically 5 iterations are enough to reach
convergence in the self-consistent calculation.

In this method we need to solve Eq.~(\ref{eq:responsep-new}) both for
the $(+\q,+\omega)$ channel and the $(-\q,-\omega)$ channel. The solutions for each channel
are evaluated separately and independently. The parallelization of the algorithm is
on the $\k$-points within each channel.
%, as well as on the planewaves defining
%each spinor wavefunction, as it is standard in Quantum ESPRESSO.

In all the calculations reported below we employed a parameter $\alpha = 500$~meV in 
Eq.~(\ref{eq:responsep-new}), which lies above the highest magnon energy calculated in 
the three examples.

\section{Results}\label{sec:benchmark}

In order to test our implementation we performed calculations on three representative
elemental transition metals, namely bcc iron, fcc nickel, and bcc chromium.
Ground-state DFT calculations were performed using Quantum ESPRESSO, within the local
density approximation for the exchange and correlation,\cite{perdew81} and using pseudopotentials 
from the repository `PSlibrary 0.3.1'.\cite{dalcorso14} 

When performing calculations of spin fluctuations, the ground state calculation and 
the solution of the Sternheimer equation must be carried out using the {\it same} sampling
of the Brillouin zone. This is necessary in order to avoid spurious symmetry-breaking
leading to the so-called gap error, that
is the presence of long-wavelength magnons with finite excitation energy.\cite{lounis11,buczek11, 
rousseau12}
In all the following calculations we employed a 50$\times$50$\times$50 grid of $\k$-points
for calculating both the ground state and the spin fluctuation spectra.

Below, when comparing with previous work, we only include the most recent theoretical studies.
A more detailed comparison between earlier theoretical results can be found in 
Refs.~\onlinecite{rousseau12, karlsson00}.

  \begin{figure*}
  \centering
  \includegraphics[width=\textwidth]{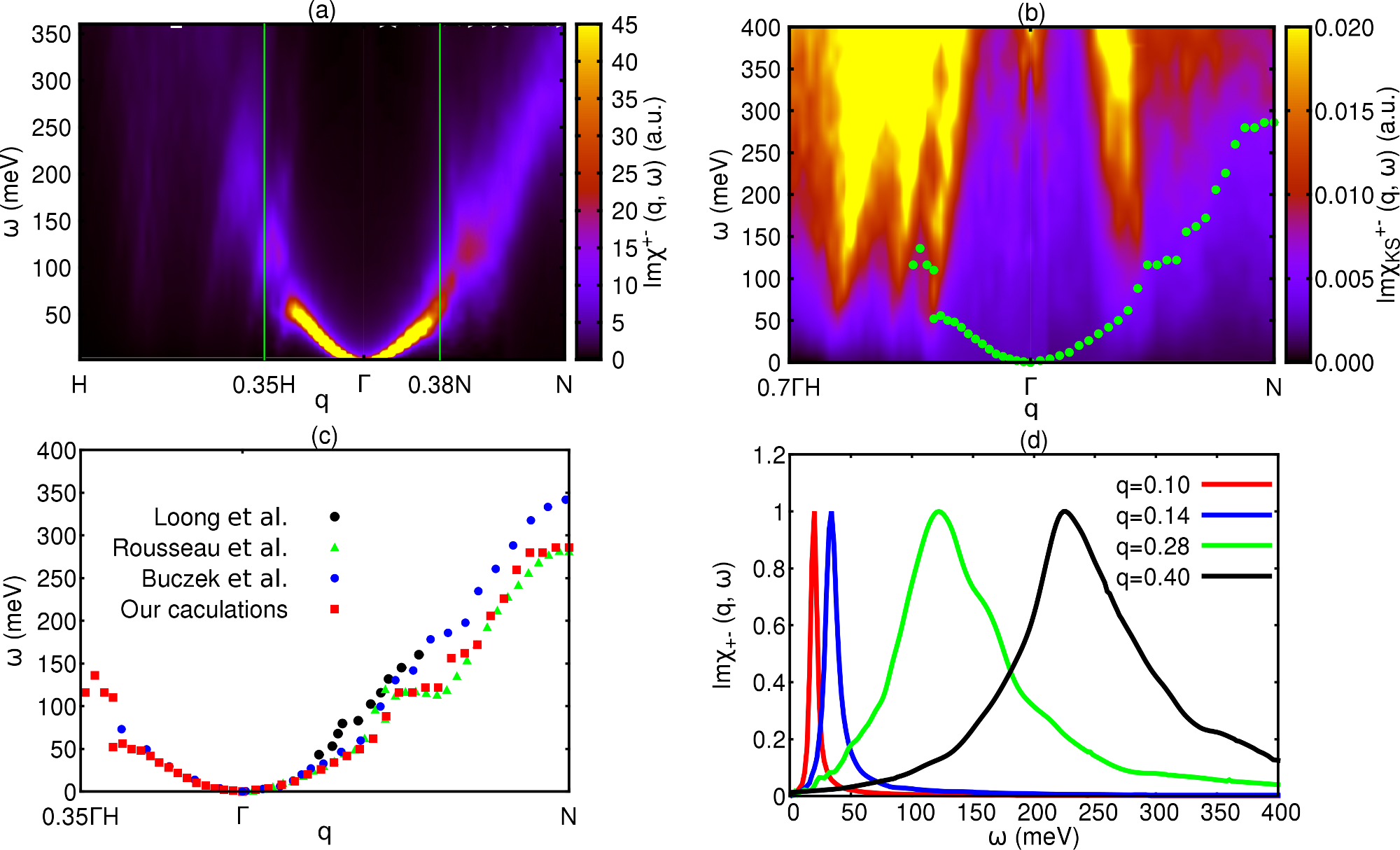}
  \caption {
  (Color online) (a) Calculated Im$\chi^{+-}(\q, \w)$ of Fe along the $\Gamma {\rm N}$ and
  $\Gamma {\rm H}$ directions. The susceptibility is given in atomic units, $1/({\rm Ry\, bohr}^3)$. (b) Calculated Im$\chi^{+-}_{\rm KS}(\q, \w)$
  along the $\Gamma {\rm N}$ and $\Gamma {\rm H}$ directions. We also show the magnon dispersion curves
  superimposed as dots. (c) Magnon bands of Fe along the $\Gamma{\rm N}$ and $\Gamma{\rm H}$ directions.
  We compare our results to the experimental data of Loong {\it et al.}\cite{loong84},
  and the calculations of Rousseau {\it et al.},\cite{rousseau12} and Buczek {\it et al.}.\cite{buczek11}
  (d) Calculated Im$\chi^{+-}(\q, \w)$ of Fe, for selected wavevectors along the $\Gamma$N direction.
  All peaks are normalized to their maximum for clarity, and the wavevectors are given as
  $\q = 2\pi/a(1,1,0)q$.}
  \label{fig:Fe}
  \end{figure*}

\subsection{BCC Iron}
\label{sec:Fe}

We performed calculations using a norm conserving pseudopotential,\cite{hamann79} using
a planewaves kinetic energy cutoff of 60~Ry. To deal with fractional occupations we used a
Gaussian smearing with a width of 10~mRy. We employed the experimental lattice parameter
$a=5.406$~bohr. Our calculations yield a ground-state magnetization of  2.16~$\mu_{\rm B}$ per
atom, and the mean Stoner splitting is 2.5~eV. We evaluated the transverse spin suceptibility 
$\chi^{+-}(\q, \w)$ for wavevectors $\q$ along the $\Gamma$N and the $\Gamma$H high-symmetry
lines [N~$ = (1/2,1/2,0)2\pi/a$, H~$ = (0,0,1)2\pi/a$]. We sampled the frequency
axis in the range 0 to 400~meV, with a spacing of 2~meV between consecutive grid points.
The broadening parameter was set to $\eta=0.1\,\w$; this choice was motivated by the
observation that a larger broadening is required to obtain converged spectra at higher frequencies.

The calculations are shown in  Fig.~\ref{fig:Fe}. In particular, Fig.~\ref{fig:Fe}(a) shows a
two-dimensional plot of Im~$\chi^{+-}$ in the $(\q, \w)$-plane. For clarity we also report
some cuts for a few selected wavevectors in Fig.~\ref{fig:Fe}(d). The maxima of the map in 
Fig.~\ref{fig:Fe}(a) are reported in the form of magnon frequency-wavevector dispersion
relations in Fig.~\ref{fig:Fe}(c). The map of Fig.~\ref{fig:Fe}(a) shows a significant 
attenuation of the magnon resonances beyond $|\q| \sim 0.38 \, \Gamma{\rm N}$, nevertheless
we can clearly recognize the magnon excitations all the way up to the Brillouin zone edge.
Along the $\Gamma$H line instead, magnon excitations are so strongly damped beyond
$|\q| \sim 0.35 \, \Gamma{\rm H}$ that no clear curve can be identified in this region.
The difference in the magnon damping patterns along the two directions is a direct consequence
of the anisotropic nature of the Stoner continuum, as it is seen in Fig.~\ref{fig:Fe}(b). Here
we see that along $\Gamma$H Stoner excitations become possible already around 100~meV, causing
the complete suppression of magnons with energies above this threshold.

  \begin{figure*}
  \centering
  \includegraphics[width=\textwidth]{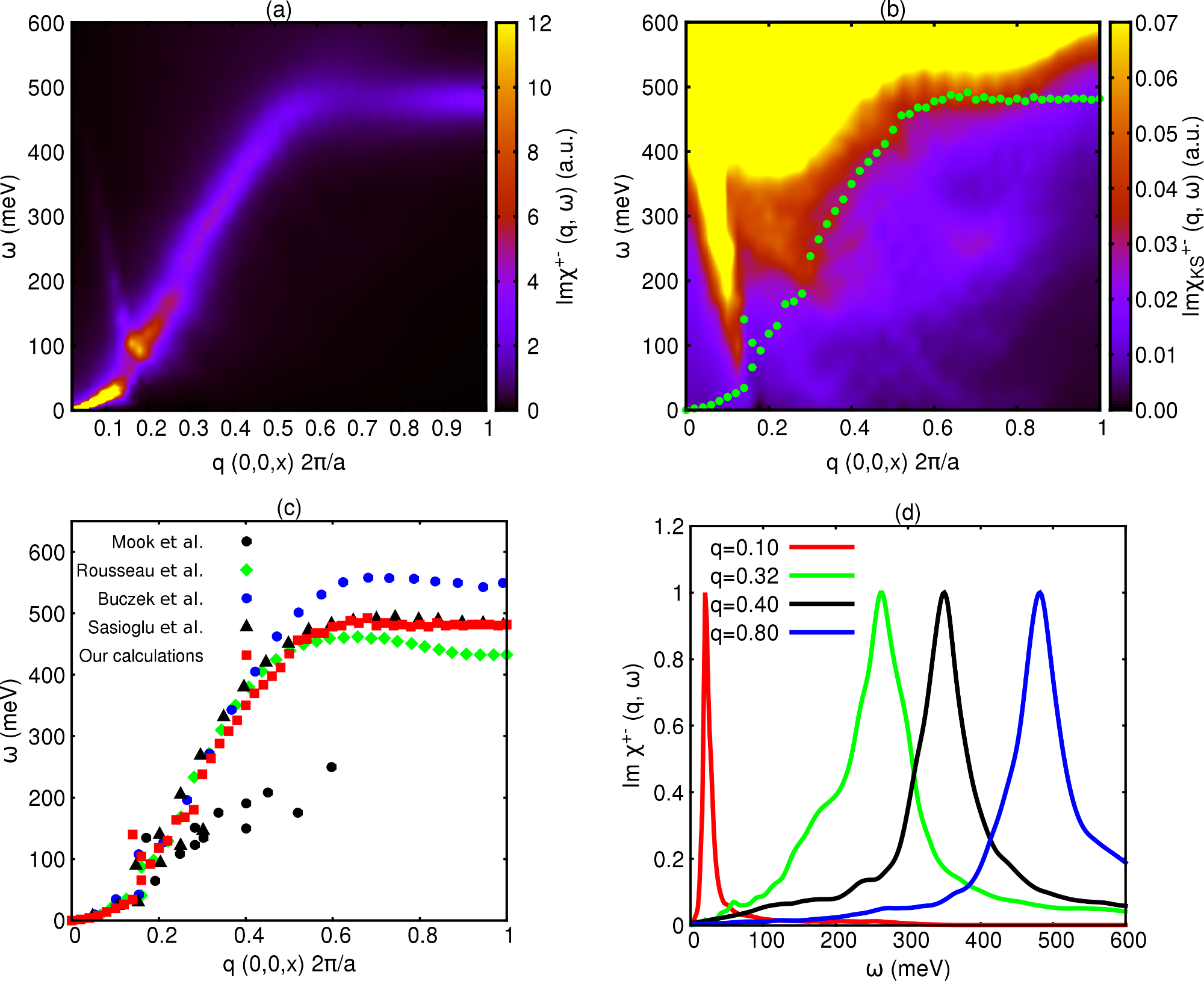}
  \caption {
  (Color online) (a) Calculated Im$\chi^{+-}(\q, \w)$ of Ni along the $\Gamma {\rm X}$ direction.
  (b) Calculated Im$\chi^{+-}_{\rm KS}(\q, \w)$, with the magnon band superimposed as dots.
  (c) Calculated magnon dispersion curve of Ni along the $\Gamma$X direction. Our results are
  compared with experimental data of Mook {\it et al.},\cite{lynn81,mook85} and with the
  calculations of Rousseau {\it et al.},\cite{rousseau12} Buczek {\it et al.},\cite{buczek11}
  and \ifmmode \mbox{\c{S}}\else \c{S}\fi{}a\ifmmode \mbox{\c{s}}\else \c{s}\fi{}\ifmmode \imath \else
  \i \fi{}o\ifmmode \breve {g}\else \u{g}\fi{}lu {\it et al.}\cite{sasioglu10}
  (d) Calculated Im$\chi^{+-}(\q, \w)$ of Ni for selected wavevectors along the
  $\Gamma$X direction. The peaks are normalized to their maximum value, and the wavevector
  is given as $\q = 2\pi/a(1, 0, 0)q$.}
  \label{fig:Ni}
  \end{figure*}

In Fig.~\ref{fig:Fe}(c) we also compare our results with previous experimental and theoretical 
work. Along the $\Gamma {\rm N}$ direction our magnon dispersion curve is in excellent agreement with 
the results of Ref.~\onlinecite{rousseau12}. This level of agreement was to be expected since
also the implementation of Ref.~\onlinecite{rousseau12} is based on the Quantum ESPRESSO package,
although these authors used the method described in Sec.~\ref{sec.dyson} instead of the Sternheimer
equation. As pointed out in Ref.~\onlinecite{rousseau12}, theoretical results tend to differ
significantly beyond $|\q| \sim 0.4\Gamma {\rm N}$. Notably, our calculations yield a 
plateau around $|\q| \sim 0.5\Gamma {\rm N}$, while Ref.~\onlinecite{buczek11}
reports a monotonic curve. We also observe that the discrepancies lie in the region where
Stoner excitations kick in. It is likely that these differences between the various approaches
arise from the difference in the underlying ground-state DFT calculations.
Along the $\Gamma {\rm H}$ direction our results are in agreement with those of Ref.~\onlinecite{buczek11},
which also find a strong suppression of spin waves in the same range of wavevectors.
Our calculations are in reasonable agreement with the low-temperature experimental data 
of Ref.~\onlinecite{loong84} (taken at 10~K). One aspect which complicates the comparison
between theory and experiment is that the measurements were not taken exactly along the
$\Gamma {\rm N}$ or $\Gamma {\rm H}$ lines. Having more information on the precise wavevector
path probed in the experiments would be useful to perform a more detailed comparison.

\subsection{FCC Nickel}\label{sec:Ni}

In this case we performed calculations using an ultrasoft pseudopotential,\cite{vanderbilt90} using
a planewaves kinetic energy cutoff of 42~Ry for the wavefunction and a cutoff of 236~Ry for the
charge density. We employed Marzari-Vanderbilt smearing\cite{marzari99} with a broadening
parameter of 10~mRy. We used the experimental lattice parameter $a=6.65$~bohr. 
We obtained a ground-state magnetization of 0.61~$\mu_{\rm B}$ per
atom and the mean Stoner splitting of 0.66~eV. We computed $\chi^{+-}(\q, \w)$ along the 
$\Gamma$X direction [X~$ = (0,0,1)2\pi/a$]. We sampled frequencies
up to 600~meV with a grid spacing of 2~meV and a broadening parameter $\eta=0.1\,\w$.

  \begin{figure*}
  \centering
  \includegraphics[width=\textwidth]{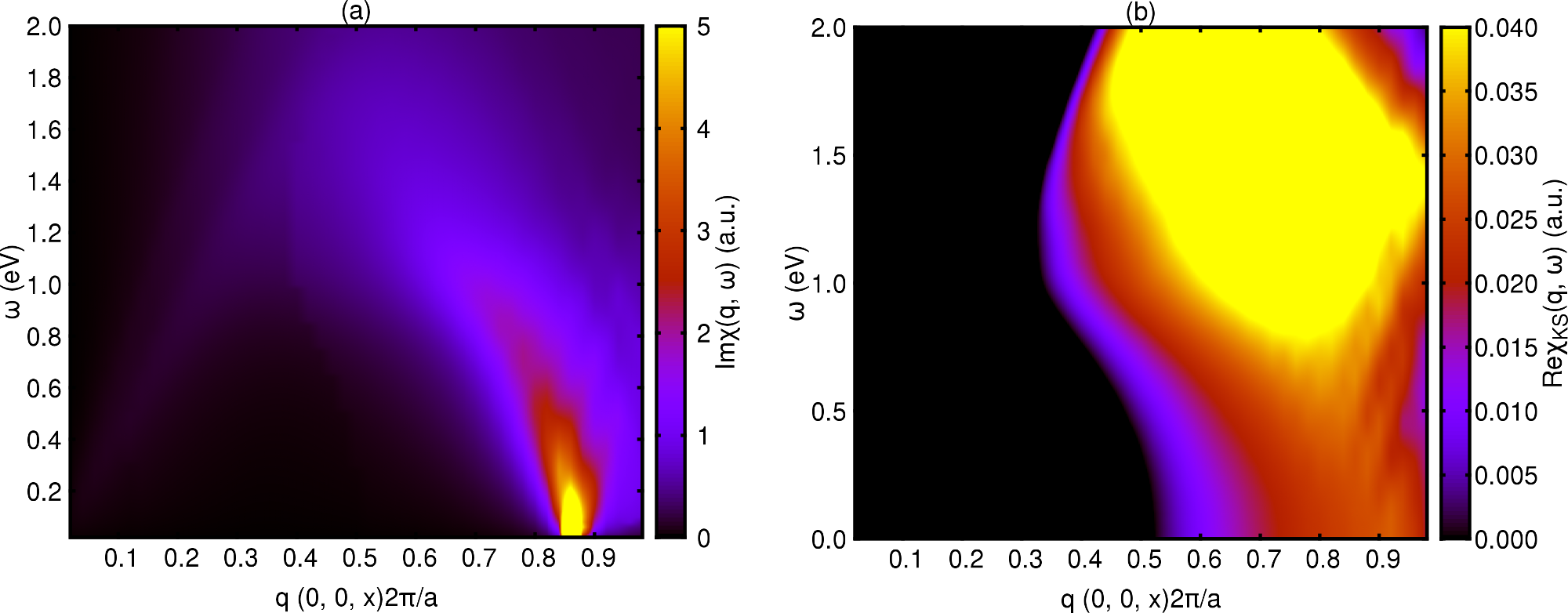}
  \caption {
  (Color online) (a) Calculated Im$\chi(\q, \w)$ of Cr along the $\Gamma$H direction.
  (b) Calculated Re$\chi_{\rm KS}(\q, \w)$ of Cr along the $\Gamma$H direction. 
	  To highlight the effect of Fermi-surface nesting, we plot Re$\chi_{\rm KS}(\q, \w)$ $-$ 0.2 and only show 
	  the positive values.}
  \label{fig:Cr_TH_2d}
  \end{figure*}

The calculated susceptibility is shown as a two-dimensional $(\q, \w)$-map in Fig.~\ref{fig:Ni}(a).
Also in this case we show representative cuts at selected wavevectors in Fig.~\ref{fig:Ni}(d).
From the map we recognize a well-defined magnon band along the $\Gamma {\rm X}$ line. The excitation
spectrum is sharp up to $|\q| \sim 0.3\Gamma {\rm X}$, and becomes broadened upon entering the Stoner
continuum beyond this point. This effect is clearly seen in Fig.~\ref{fig:Ni}(b), where we superimpose
the spin wave dispersion curve to the Stoner spectrum.
Our calculated magnon band, as extracted from the maxima of Im$\chi^{+-}(\q, \w)$, is compared to 
previous calculations and experiments in Fig.~\ref{fig:Ni}(c). Here we observe that theoretical
data agree well for small wavevectors, but start deviating significantly from each other near
the zone boundary. which is also observed in Ref.~\onlinecite{rousseau12}. 
We tentatively assign these deviations to the different choices employed to
describe the ground-state electronic structure.

When comparing with experiment we see that our calculations, as well as previous theoretical
data, consistently overestimate the experiments by up to a factor of two. This phenomenon
is well understood, and is related to the fact that the Stoner splitting of fcc Ni in standard
DFT is approximately twice as large as in experiment (0.66~eV vs. 0.3~eV,
respectively).\cite{raue1984, eastman80, sasioglu10} 
It was shown that the experimental magnon spectra can be reasonably reproduced by manually reducing 
the LSDA exchange splitting by one-half.\cite{karlsson00,sasioglu10}

\subsection{BCC Chromium}

For chromium we used a norm-conserving pseudopotential with a cutoff of 60~Ry, and a Gaussian
smearing of 10~mRy. We set the lattice parameter to the experimental value $a=5.5$~bohr.
We evaluated the susceptibility along the $\Gamma$H line, and we sampled frequencies using
a grid spacing of 20~meV, up to a maximum value of 2~eV. In this case we used a constant
broadening parameter $\eta = 100$~meV.

Experiments indicate that the ground-state electronic structure of Cr is a spin-density wave (SDW),
with an incommensurate wave vector $\q_{\rm SDW}$ $\simeq (0, 0, 0.95) 2\pi/a$.\cite{fincher79}
This SDW was attributed to the presence of pronouced Fermi-surface nesting in paramagnetic 
Cr,\cite{overhauser62, lomer62,fawcett88} a scenario which was later confirmed by explicit
DFT calculations.\cite{hafner02}

The calculated spin susceptibility of paramagnetic Cr is shown in Fig.~\ref{fig:Cr_TH_2d}(a).
Here we see a clear peak for wavevectors around $(0, 0, 0.86)2\pi/a$, extending down to
zero excitation energy. The presence of zero-energy excitations at finite wavevector is
the signature of a SDW, i.e. a frozen spin-wave, analogous to charge-density waves observed
in the presence of soft phonons. The role of Fermi-surface nesting can be investigated by
inspecting the nesting function, that is  Re$\chi_{\rm KS}(\q, \w)$, as shown in Fig.~\ref{fig:Cr_TH_2d}(b).
For $\w=0$ the nesting function reaches a maximum at $\q_{\rm nest} \sim (0, 0, 0.92)2\pi/a$, 
which is close to our calculated SDW vector. This result confirm that nesting is at the
origin of the SDW in chromium.

Our present results are in good agreement with previous calculations,\cite{savrasov98} 
although we obtain a slightly shorter SDW vector (0.86$\cdot2\pi/a$ vs. 0.92$\cdot2\pi/a$). We 
assign this difference to the fact that we used the LDA functional, while 
in Ref.~\onlinecite{savrasov98} the PBE functional was employed.\cite{perdew96}

\section{Conclusions}\label{sec:conclusion}

In conclusion, we described an implementation of time-dependent density-functional perturnation
theory for spin fluctuations, based on planewaves and pesudopotentials, built around the linear
response modules of the Quantum ESPRESSO package. In the present approach we calculate the
macroscopic spin susceptibility $\chi(\q, \w)$ via a self-consistent solution of the time-dependent
Sternheimer equation. The main advantage of this formulation is that it avoids altogether the
need for evaluating unoccupied Kohn-Sham states. 

We demonstrated the methodology by calculating spin wave spectra of bcc Fe, fcc Ni, and bcc Cr
along several high-symmetry directions in the Brillouin zone, and we rationalized the suppression
of magnon excitations in terms of Landau damping when the magnon energy resonates with the Stoner 
continuum. In the case of Fe and Ni, our calculated magnon dispersions relations are in good
agreement with previous theoretical results near the zone center, but we see some significant
deviations closer to the zone boundaries. These deviations likely result from the
underlying DFT description of the ground-state electronic structure. In the case of Fe, our
calculations are in reasonable agreement with experiment, while in the case of Ni the calculated
magnon energies are too large by a factor of two. This discrepancy is consistent 
with previous {\it ab initio} calculations, and is attributed to the fact
that standard DFT yields too large a Stoner splitting for this metal as compared to experiment.
In the case of Cr, we demonstrated that the calculation of spin wave spectra can be a very powerful
tool to identify SDW phases, and we obtained good agreement with previous theory and with experiment.

Looking forward, we see two main avenues for future development. Firstly, it would be desirable
to extend the present formalism and implementation to Hubbard-corrected DFT. The capability of
calculating the spin suceptibility within DFT+U would make it possible to explore many interesting
correlated electron system, including for example the copper oxides high-temperature superconductors. 
Secondly, it would be
important to incorporate support for spin-orbit coupling in the methodology. This further development
will enable calculations of spin wave spectra on systems with strong magnetic anisotropy, for example
multiferroic oxides. Apart from these desirable improvements, we believe that
our current implementation provides an important new addition to state-of-the-art techniques
for investigating spin fluctuations, magnon dispersions, and spin density waves in several important 
problems of condensed matter physics, and will serve as a starting point to
explore incommensurate magnetic excitations systematically, without performing supercell calculations.

\vspace{10pt}{\it Note.} We are aware of a closely related work, 
where the authors used TD-DFPT to calculate magnon dispersion relations. \cite{gorni_thesis} The present 
results are consistent with those of Ref.~\onlinecite{gorni_thesis}.

\begin{acknowledgments}
This work was funded by EPSRC grant No. EP/M020517/1, entitled ``Oxford Quantum Materials Platform Grant"
and the Leverhulme Trust (Grant RL-2012-001). The authors acknowledge the use of the
University of Oxford Advanced Research Computing (ARC) facility (http://dx.doi.org/10.5281/zenodo.22558) 
and the ARCHER UK National Supercomputing Service.
\end{acknowledgments}

%\bibliography{references}

%\end{document}

\end{document}